\documentclass{aa}
\usepackage{graphicx}

\begin{document}
\title{Obscured clusters.\,I. GLIMPSE\,30 -- Young Milky Way Star Cluster Hosting
Wolf-Rayet Stars\thanks{Based on observations collected with the New
Technology Telescope of the ESO within observing program 77.D-0089}}

\author{R. Kurtev
   \inst{1}\fnmsep\thanks{``Centro de Astrof\'isica de Valpara\'so''. Visiting astronomer ESO La Silla
Paranal Observatory}
   \and
   J. Borissova\inst{1}
   \and
   L. Georgiev\inst{2}
   \and
   S. Ortolani\inst{3}
   \and
   V.D. Ivanov\inst{4}         }

\offprints{R. Kurtev}

\institute{
  Departamento de Fis\'ica y Astronom\'ia, Facultad de
  Ciencias, Universidad de Valpara\'iso, Av. Gran Breta\~na 644,
  Playa Ancha, Casilla 5030, Valpara\'iso, Chile \\
    \email{radostin.kurtev@uv.cl; jura.borissova@uv.cl}
\and
  Instituto de Astronomia, Universidad Nacional Aut\'onoma de M\'exico,
  Apartado Postal 70-254, CD Universitaria, CP 04510 Mexico DF, Mexico,  \\
    \email{georgiev@astroscu.unam.mx}
\and
  Universit\'a di Padova, Dipartimento di Astronomia, Vicolo
  dell'Osservatorio 5, I-35122 Padova, Italy
    \email{sergio.ortolani@unipd.it}
    \and
  European Southern Observatory, Ave. Alonso de Cordova 3107,
  Casilla 19, Santiago 19001, Chile \\
    \email{vivanov@eso.org}
}

\date{Received; accepted}

\abstract
{Young massive clusters are usually deeply embedded in dust and gas.
They represent perfect astrophysical laboratories for study of massive
stars. Clusters with Wolf-Rayet (WR) stars are of special
importance, since this enables us to study a coeval WR population
at a uniform metallicity and known age.}
{We started a long-term project to search the inner Milky Way for hidden star
clusters and to study them in details. GLIMPSE\,30 (G30) is one of them.
The cluster is situated near the Galactic plane ($l$=298$\fdg$756,
$b$=$-$0$\fdg$408) and we aimed to determine its physical parameters
and to investigate its high-mass stellar content and especially WR stars.}
{Our analysis is based on SOFI/NTT $J_{\rm S}HK_{\rm S}$ imaging and
low resolution (R$\sim$2000) spectroscopy  of the brightest cluster
members in the K atmospheric window. For the age determination we applied
isochrone fits for MS and Pre-MS stars. We derived stellar parameters
of the WR stars candidates using a full nonLTE modeling of the
observed spectra.}
{Using a variety of techniques we found that G30 is very young cluster, with age $t$ $\approx$ 4\,Myr.
The cluster is located in Carina spiral arm, it is deeply embedded in
dust and suffers reddening of $A_{V}$\,$\sim$\,10.5$\pm$1.1\,mag. The
distance to the object is d=7.2$\pm$0.9\,kpc. The mass of the cluster
members down to 2.35\,$\cal M_\odot$ is $\sim$\,1600\,$\cal M_\odot$.
Cluster's MF for the mass range of 5.6 to 31.6\,${\cal M_\odot}$ shows a
slope of $\Gamma$=$-1.01$$\pm$$0.03$. The total mass of the cluster obtained
by this MF down to $1$\,$\cal M_\odot$ is about $3\,\times\,10^3$\,$\cal M_\odot$.
The spectral analysis and the models allow us to conclude that in G30 are
at least one Ofpe/WN and two WR stars. The WR stars are of WN6-7 hydrogen
rich type with progenitor masses more than 60\,$\cal M_\odot$.}
{G30 is a new member of the exquisite
family of young Galactic clusters, hosting WR stars. It is a factor
of two to three less massive than some of the youngest super-massive star clusters
like Arches, Quintuplet and Central cluster and is their smaller analog.}


\authorrunning{R.\,Kurtev et al.}
\titlerunning{GLIMPSE\,30 -- Young Milky Way Star Cluster Hosting
Wolf-Rayet Stars}

\maketitle
%

\section{Introduction}

Stars rarely form in isolation. In fact, it is
well known that more of the stars in our Galaxy,
and in nearby galaxies, are born in groups ranging from small
associations and open clusters, compact young massive clusters
to old globulars. Young clusters are often difficult to find because they
can be heavily embedded in dust, making them visible only in the infrared.
We embarked on a long-term project to search the inner Milky Way for hidden star
clusters and to study them in details (Borissova et al. 2003, 2005,
2006; Ivanov et al. 2002, 2005). This project was based on the 2MASS (Skrutskie et al.
2006), taking advantages of the reduced extinction in the near-IR.
Recent advances in mid-IR instrumentation made it possible to carry out
all-sky IR surveys in this spectral region too. The recent {\it Spitzer Space
Telescope} Galactic Legacy Infrared Mid-Plane Survey Extraordinaire
(GLIMPSE, Benjamin et al. 2003) offers an excellent opportunity
to carry out a deep census of such objects. GLIMPSE is an excellent tool
for finding obscured clusters in the Galactic disk because the extinction
in the mid-IR is a factor of 2-5 lower than in the near-IR.
Recently, Mercer et al. (2005) implemented a comprehensive search for
clusters in the mid-IR. They used the point source catalog of the GLIMPSE
and reported 92 cluster candidates. In our project we continued the
investigation of some of these objects using near-IR imaging and
low resolution IR spectroscopy.

Massive stars themselves play an important role in the ecology of
galaxies, providing a major source of ionizing UV radiation,
mechanical energy and chemical enrichment. Wolf-Rayet (WR) stars represent an
evolved phase of the most massive stellar population, and are characterized by
high mass loss rates from fast and dense winds. Their short lifetimes and
high luminosities make them excellent tracers of active recent star formation.
However, serious gaps in our understanding of massive stars exist because of
the rapid evolution and rarity. WRs in clusters are particularly interesting
because this enables us to study a coeval population with uniform metallicity.
There are about three hundred known WR stars in our Galaxy (van der Hucht 2006).
At the same time from thousands known open clusters and stellar associations
small number contain WRs and only a few of them host more than one such star.
This is because a typical cluster ($\sim$10$^3$\,$\cal M_\odot$) contains a
limited number of stars more massive than $\sim$30-35\,$\cal M_\odot$
that can evolve to the WR phase. Therefore, most cluster WRs are
concentrated in the high-mass clusters, i.e. Arches, Quintuplet, Westerlund\,1.

\begin{figure}
\centering
\includegraphics[width=\columnwidth]{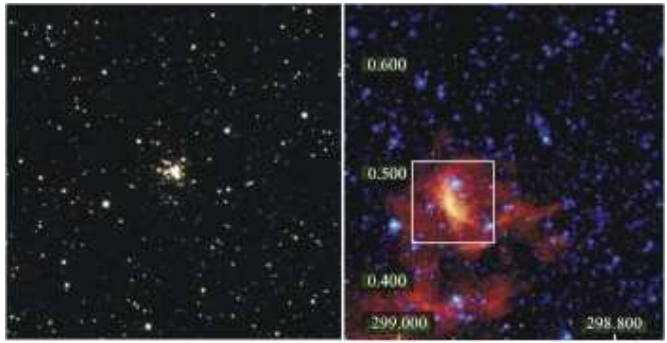}
\caption{Pseudo-true color images of the G30. The left panel image
is composed from $J_{\rm S}$ -- blue, $H$ -- green and $K_{\rm S}$ --
red SofI/NTT images covering 4\farcm92$\times$4\farcm92 centered at
the cluster. The right panel is composed from 3.6\,$\mu$m -- blue,
5.8\,$\mu$m -- green and 8.0\,$\mu$m -- red {\it Spitzer Space
Telescope} images. The SofI field of view from the left panel is
marked with a square. The Galactic coordinates $l$ and $b$ are shown
on this right panel.}
\label{fig1}
\end{figure}

Here we report our first results for GLIMPSE\,30 (hereafter G30;
Figure\,\ref{fig1}) -- compact young cluster located near the Galactic
plane ($l=298\fdg756, b=-0\fdg408$) containing many massive stars
including at least one Of and three Wolf-Rayet members. The
presence of at least three WRs in G30 puts this cluster into the
family of WR rich clusters. The cluster membership provides excellent
observational constraints upon the ages and initial masses of this type
of stars. In this paper we present the main results of our
photometric and spectral analysis of the cluster and the newly
discovered WRs.

\section{Observations and Data Reduction}

All observations were obtained with SofI/NTT (Son of ISAAC;  Moorwood,
Cuby \& Lidman 1998). The instrument is equipped with Hawaii HgCdTe
1024$\times$1024 detector, with pixel scale of 0.288\,arcsec\,px$^{-1}$.
For the spectroscopy we used 1\,arcsec slit and the medium-resolution
grism. The seeing for all observations was 1-1.5\,arcsec
and the sky was photometric.

Deep $J_{\rm S}HK_{\rm S}$ imaging of G30 were carried out on April 15,
2006. We took 16 images in each filter in jittering mode with 3\,arcmin
jitter box size to ensure that there is no overlapping of the cluster
on different images. Each individual image was the average of
3$\times$20\,sec frames in $J_{\rm S}$, 6$\times$10\,sec frames in
$H$, and 10$\times$6\,sec frames in $K_{\rm S}$. The total integration 
time was 16\,min in each filter. To obtain photometry of
the brightest cluster members we took additional shallow images on
August 10, 2006, using the same jittering pattern but shorter
integrations: 5$\times$1.182\,sec. The total integration time was 1.58\,min 
in each filter. The data reduction included: sky-subtraction, flat-fielding,
image alignment and combination into a single final image for each filter.
A 3-color composite image of G30 is shown on Figure~\ref{fig1}.

The stellar photometry of the final images was carried out with
{\sc ALLSTAR} in {\sc DAOPHOT II} (Stetson 1987). The typical
photometric errors vary from 0.01\,mag for stars with
$K_{\rm S}$ $\sim$10\,mag to 0.10\,mag for $K_{\rm S}$ $\sim$18\,mag
and 0.15\,mag for $K_{\rm S}$ $\sim$19\,mag. The photometric
calibration was performed by comparing our instrumental magnitudes
with the 2\,MASS measurements of about 1330 stars, covering the color
range 0.0$\leq$$J_{\rm S}$$-$$K_{\rm S}$$\leq$3.0\,mag and magnitude range
10$\leq$$K_{\rm S}$$\leq$14.5\,mag. A least squares fit of the instrumental
$jhk$ magnitudes to the standard 2MASS system gave the following relations:

\begin{eqnarray*}
&&(J_{\rm S}-K_{\rm S})-(j-k)= -1.302_{(\pm0.005)} \\
&&J_{\rm S}-j = 0.005_{(\pm 0.007)}\times (J_{\rm S}-K_{\rm S})-2.435_{(\pm0.003)} \\
&&H-h = -0.021_{(\pm 0.008)}\times (J_{\rm S}-K_{\rm S})-2.457_{(\pm0.003)} \\
&&K_{\rm S}-k = -0.004_{(\pm 0.006)}\times (J_{\rm S}-K_{\rm S})-1.134_{(\pm0.005)}
\end{eqnarray*}

The final photometry list contains equatorial coordinates, and
$J_{\rm S}HK_{\rm S}$ magnitudes of 7469 stars with photometric errors
less than 0.15\,mag. Artificial stars tests show that the 80\% completeness
limit of the photometry is at $J_{\rm S}$=19.1 and $K_{\rm S}$=17.4.

The spectra were obtained on Apr 14, 2006. They cover the region from
$\sim$2.00 to $\sim$2.35\,$\mu$m. The slit was aligned on a sample of
bright cluster members and the telescope was nodded along the slit
between the exposures. We obtained 16 images of 150\,sec each, on one single
slit position. In total, spectra of 8 stars were extracted from
the data. To correct for the telluric absorption we observed the star
HIP59642 (HD106290) -- solar near-analog of spectral class G1V.
The reduction process included: sky subtraction, flat fielding,
geometric distortion correction, image alignment and combination into
a final 2-dimensional spectrum. Next, we extracted 1-dimensional
spectra, wavelength calibrated them, and corrected them for the
telluric absorption by multiplying with the telluric standard. After
this the target spectra were multiplied by a solar spectrum to remove
the artificial emission lines due to the intrinsic absorption features
in the spectra of the standard (see Maiolino, Rieke \& Rieke 1996).

\section{$J_{\rm S}HK_{\rm S}$ Color-Magnitude and Two-Color Diagrams}

\subsection{Field-star decontamination\label{decontam}}

The cluster G30 is located extremely close to the Galactic plane
and therefore, in a crowded stellar field. However, it is rather
concentrated and occupies small area with diameter $\sim$1.3\,arcmin,
always near the center of the SofI field of view. This allows us to
define and subtract reliable the fore- and background field star
contamination.

\begin{figure}
\centering
\includegraphics[width=\columnwidth]{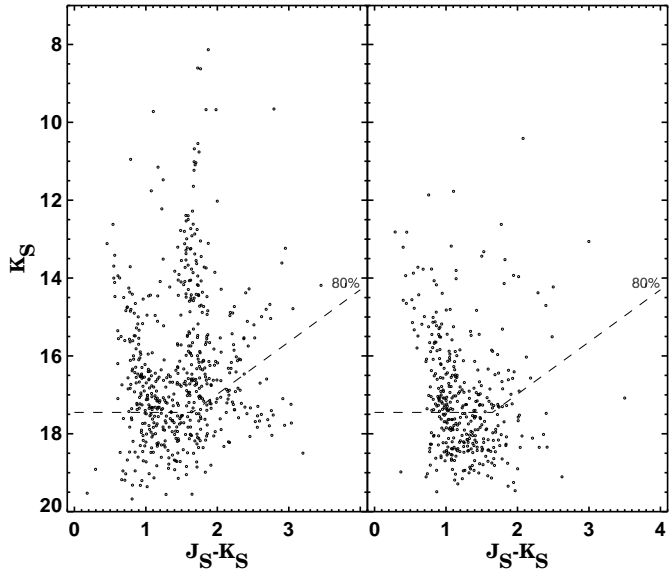}
\caption{($J_{\rm S}$$-$$K_{\rm S}$, $K_{\rm S}$) color-magnitude
diagrams of the stars within the circle with radius R=43.2\,arcsec
(R=150\,px) around G30 (left panel) and a comparison field (right
panel). See Section~\ref{decontam} for details.}
\label{fig2}
\end{figure}

Photometric and spatial criteria were used to select probable
cluster members. First, only the stars in a circle with radius
R=150\,px ($\sim$43\,arcsec), similar to the apparent cluster size,
and centered on the cluster were chosen. This limits the candidate
members to 592 stars. The color-magnitude diagram (CMD) of this
sample is shown on the left panel of Figure\,\ref{fig2}. The locus
of the cluster Main Sequence (MS) stars can be easily seen at
1.5$<$(J$_{\rm S}$\,$-$\,K$_{\rm S}$)$<$2\,mag. Some fore- and
background stars are also present on this diagram. In order to
define their locus, the CMD of the stars falling into a circle with
the same radius but centered 100\,arcsec to the North-East from the
cluster's center is shown on the right panel of Figure\,\ref{fig2}.
The contamination become significant only at $K>$17\,mag while the
rest of the cluster locus is almost entirely devoted of field stars.

\begin{figure}
\centering
\includegraphics[width=\columnwidth]{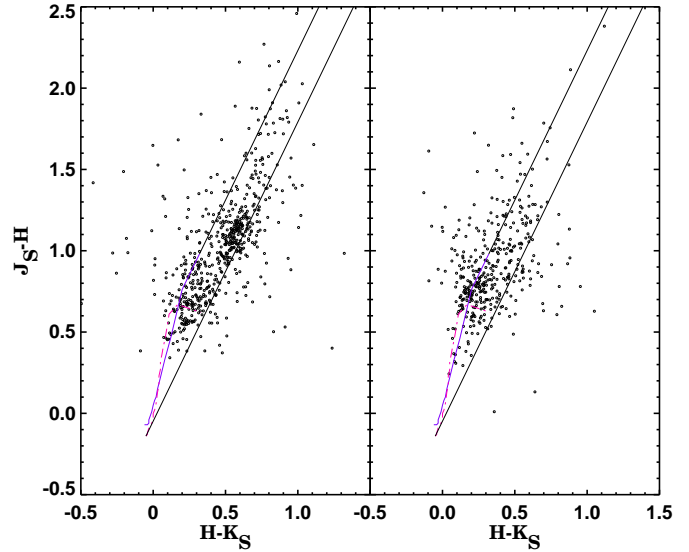}
\caption{($H$$-$$K_{\rm S}$, $J_{\rm S}$$-$$H$) two-color diagrams
of the stars within the circle with radius R=43.2\,arcsec (R=150\,px)
around G30 (left panel) and the comparison field (right panel). The
continuous and dotted lines represent the sequence of the
zero-reddening stars of luminosity classes I (Koornneef 1983) and V
(Schmidt-Kaler 1982), respectively. Reddening vectors for O5V and M5I
stars are also shown.}
\label{fig3}
\end{figure}

The majority of the cluster candidate members occupy a well-defined locus
on the two-color diagram (TCD) as well, at $H$$-$$K_{\rm S}$$\sim$0.6\,mag
and $J_{\rm S}$\,$-$\,$H$$\sim$1.1\,mag. Figure\,\ref{fig3} shows that there
is only small number of field stars falling in the cluster locus on the
right panel of the same figure where the TCD of the comparison field
(defined above) is shown.

The decontamination procedure was based on the color and magnitude
distributions of the field stars in five stellar fields around G30, to
improve the field star statistics. Each field covers the same area as
the cluster. Their combined CMD was divided into bins and the number
of stars in each bin was divided by five to normalize the combined
field area to the cluster area. Then, we randomly subtracted from the
corresponding bin of the cluster's CMD as many stars as were present
in the corresponding field CMD bin. The bin sizes varied depending on
the number of stars per bin. In other words, we merged nearby bins, if
they contained less than 2 stars. The total number of stars subtracted
throughout the decontamination was 191.

\subsection{Analysis of the decontaminated diagrams}

\begin{figure*}
\centering
\includegraphics[width=\textwidth]{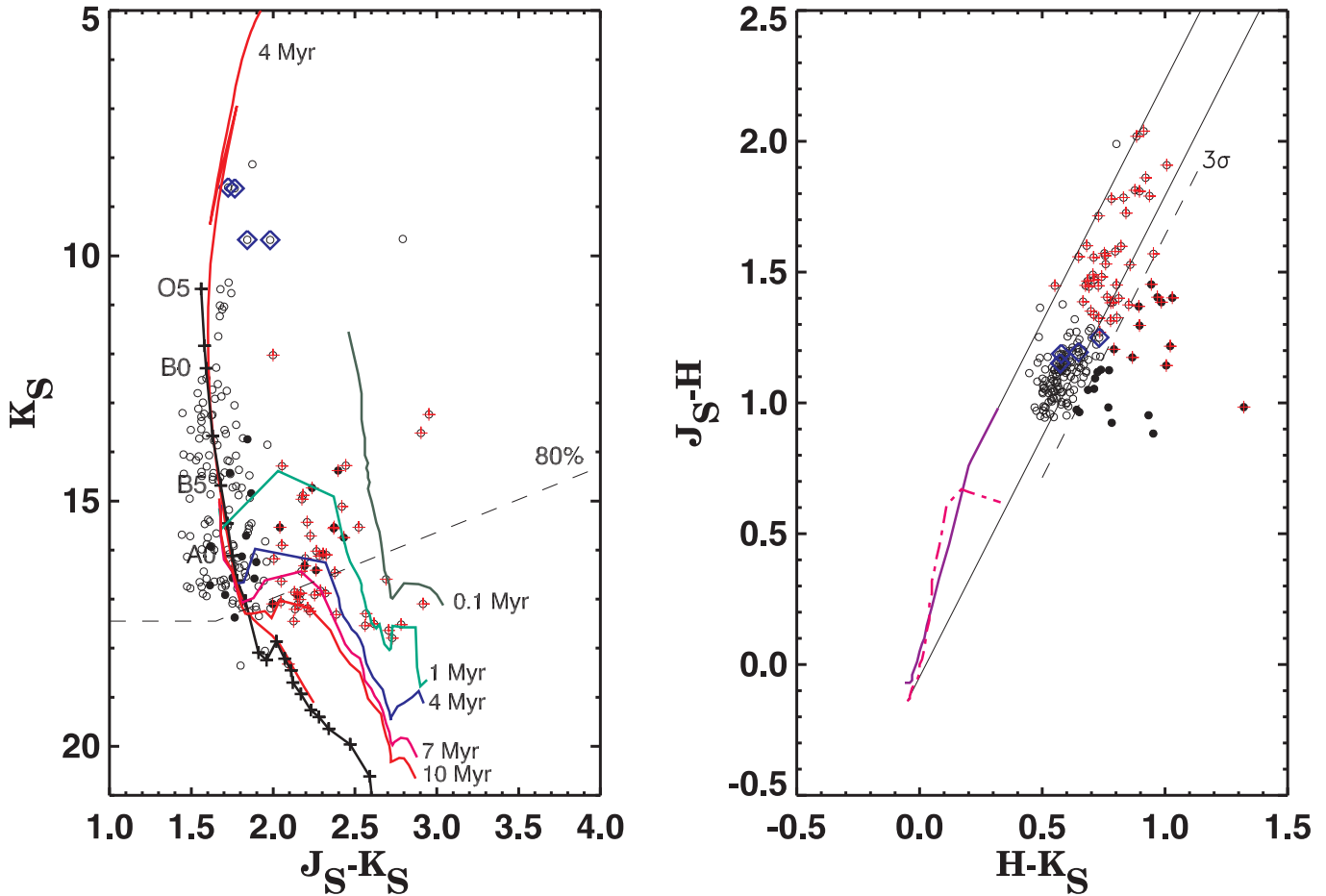}
\caption{
{\it Left:} Decontaminated ($J$$-$$K_{\rm S}$, $K_{\rm S}$) color-magnitude
diagram. An isochrone for 4\,Myr from Girardi et al. (2002) and PMS isochrones
for various ages and solar metallicity from (Siess, Dufour \& Forestini 2000) 
are shown. The 80\% completeness limit of
the photometry is marked with a dashed line. Solid dots indicate stars with
color excess, as defined on the right panel.
{\it Right:} Decontaminated ($H$$-$$K_{\rm S}$, $J_{\rm S}$$-$$H$) two-color
diagram. The continuous and dotted lines represent the sequence of the
zero-reddening stars of luminosity classes I (Koornneef 1983) and V
(Schmidt-Kaler 1982), respectively. Reddening vectors for O5V and M5I stars
are also shown. The probable PMS stars are marked with crosses. The
$3\,\sigma$-limit line  of the reddening vector for OV star is drawn
$\approx$0.12\,mag to the right of the reddening vector. The stars redder than
this line show color excess and are marked with solid circles.
}
\label{fig4}
\end{figure*}

The decontaminated CMD and TCD of the cluster are shown in Figure\,\ref{fig4}.
The CMD morphology presents an extended MS reaching down to $\sim$17\,mag.
The majority of these stars form well-defined compact sequence on the TCD
around $H$$-$$K_{\rm S}$$\sim$0.6\,mag. The spectroscopically confirmed Of/WN
and WR stars (marked with diamonds) are among the brightest stars in the
field and they are located to the right of MS, as expected for evolved
objects. There is a hint for deviation of the most massive stars from the MS
at 10$<$$K_{\rm S}$$<$12\,mag.

The CMD shows a population of 45 likely pre-main sequence stars (PMS; marked
with crossed circles) at 1$<$$K_{\rm S}$$<$17\,mag and
2$<$$J_{\rm S}$$-$$J_{\rm S}$$<$2.5\,mag. We can not exclude the possibility
that some of them might be unresolved binaries or MS stars subjected to higher
differential reddening. To constrain the age and masses of these stars we
applied isochrone fit with PMS tracks of different ages: 0.1, 1, 4, 7, and
10\,Myr from Siess, Dufour \& Forestini (2000) shifted with the obtained
distance modulus and reddening to the cluster (see Section\,\ref{param} for
details). Considering the photometric errors and the crowding effects we reach
tentative conclusion that the majority of the PMS candidates have ages between
1 and 10\,Myr.

Non-negligible number of objects (23) fall to the right of the reddening vector
for the reddest stars on TCD. Most likely they have excess due to circumstellar
envelopes and/or discs. We marked with solid dots those stars that are more than
3$\sigma$$\sim$0.12\,mag redder than the reddening vector for the reddest MS
stars.

\section{Wolf-Rayet Stars}

\begin{figure}
\centering
\includegraphics[width=\columnwidth]{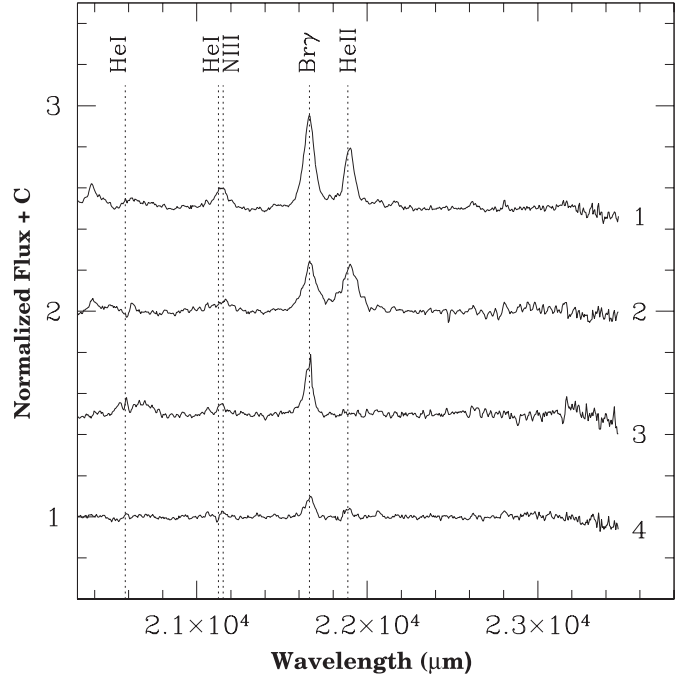}
\caption{Infrared spectra of the G30 stars. The spectra are continuum
normalized and shifted vertically by 0.5 for display purposes. The
S/N at $\lambda$=2.07$\mu$m for the stars 1 to 4 is approximately 90,
80, 80, and 60, respectively.}
\label{fig5}
\end{figure}

The Ofpe/WNL and WR members of G30 are marked with diamonds in
Figure~\ref{fig4}). Their coordinates and infrared broadband
properties are listed in Table\,\ref{tab1}, and their spectra are
plotted in Figure~\ref{fig5}. The S/N at $\lambda$=2.07$\mu$m
for the stars 1 to 4 is 90, 80, 80, and 60, respectively.

\begin{table}
\caption{Of and WR stars in G30.}
\begin{center}
\begin{tabular}{cccccc}
\hline\hline
& & & & &  \\[-8pt]
\multicolumn{1}{c}{ID} &
\multicolumn{1}{c}{$\alpha_{2000}$} &
\multicolumn{1}{c}{$\delta_{2000}$} &
\multicolumn{1}{c}{$J_{\rm S}$$-$$H$} &
\multicolumn{1}{c}{$H$$-$$K_{\rm S}$} &
\multicolumn{1}{c}{$K_{\rm S}$} \\
& & & & &  \\[-10pt]
\hline
& & & & &  \\[-7pt]
1  & 12:14:33.91 & $-$62:58:48.7 & 1.250 & 0.732 & 9.676 \\
& & & & &  \\[-10pt]
2  & 12:14:33.09 & $-$62:58:51.0 & 1.194 & 0.648 & 9.674 \\
& & & & &  \\[-10pt]
3  & 12:14:31.76 & $-$62:58:51.9 & 1.188 & 0.577 & 8.625 \\
& & & & &  \\[-10pt]
4  & 12:14:31.54 & $-$62:58:54.3 & 1.152 & 0.573 & 8.605 \\
& & & & &  \\[-10pt]
\hline
\end{tabular}
\end{center}
\label{tab1}
\end{table}

The equivalent widths of the emission lines were measured from the
normalized spectra. To quantify the subtypes of WR stars, we compared
our data with the near-IR spectral atlases of Crowther \& Smith
(1996) and Figer, McLean \& Najaro (1997), following the spectral
classification scheme of Crowther et al. (2006). The results are
summarized in Table~\ref{tab2}.

Our analysis shows that star No.\,1 is WN7, based on the presence of
strong N\,{\sc III} at 2.116\,$\mu$m. The He\,{\sc II} line at
2.189\,$\mu$m is weaker than Br${\gamma}$, but He\,I lines are also
weak.

Star No.\,2 is WN6 because the observed ratio He\,{\sc II}
2.189\,$\mu$m\,/\,Br${\gamma}$\,$>$\,1 shows lack of hydrogen.
The spectrum is similar to that of WR136 = HD192163 (van der Hucht
2001).

Star No.\,3 is particularly interesting. The presence of
Br${\gamma}$ emission, the weak He\,I lines and the lack of He\,{\sc II}
lines neither in absorption nor in emission suggest that it is
Ofpe/WNL. The spectrum resembles these of HD\,47129 (O8p),
HD\,148937 (O6fp), HD\,152408 (O8Iafpe), and HD\,269582 (Ofpe/WN9).
There are only several known stars of this type in the Galaxy.
Interestingly, Ofpe stars also reside in clusters: in Quintuplet
(Figer, McLean \& Morris 1999) and in the Galactic Center cluster
(Cotera et al. 1999). This warrants further study of the star.

\begin{figure}
\centering
\includegraphics[width=\columnwidth]{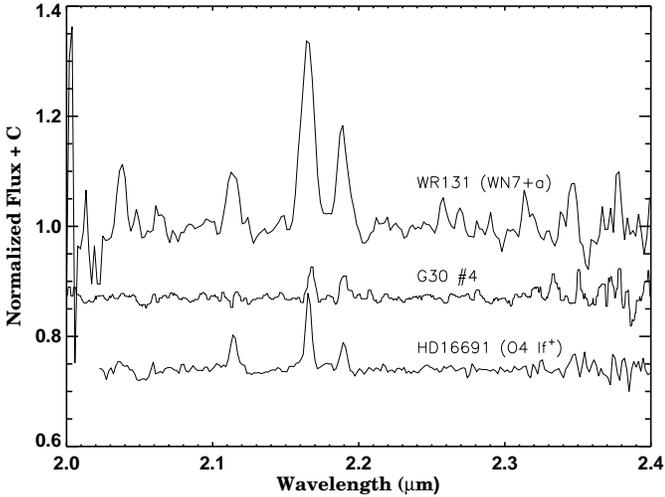}
\caption{Comparison of the spectrum of star \#4 with template
spectra of WR131 (WN+a) and HD16691 (O4\,If$^+$) from Figer,
McLean \& Najaro (1997) and Hanson, Conti \& Rieke (1996).}
\label{fig6}
\end{figure}

Star No.\,4 shows strong He\,{\sc II} line (as strong as $\rm Br{\gamma}$),
but it also has He\,I and no N\,{\sc III} line. The measured line
intensities (Table~\ref{tab2}) suggest a tentative WN7-8 classification.
Unfortunately, the star was located at the edge of the slit and its
spectrum has the lowest S/N ratio of all four WR and O star candidates.
We compared this spectrum with the template spectra of WN and O stars
given in Conti et al. (1995), Hanson, Conti \& Rieke (1996), Figer,
McLean \& Najaro (1997) and Crowther et al. (2006), smoothing the
templates to the resolution of our data. The closest matches are with
WR131 (WN+a) and HD16691 (O4\,If$^+$), shown in Figure\,\ref{fig6}.
Note that the intensities of He\,{\sc II} and Br${\gamma}$ are
higher in the template spectra and they have much more reliable 
He\,{\sc I} + N\,{\sc III} line. This comparison also shows similarities 
with WN7 class but the He{\sc II} emission suggest early or mid-O star.

Another four stars fall into the slit by serendipitously but their S/N is
too low for quantitative analysis and they show no strong emission lines.

\begin{table*}
\caption{Spectral classification of Of and WR stars in G30. For every
object the equivalent widths (first line) and FWHM (second line) in
\AA\ for prominent near-IR lines in the spectrum are presented. The
central wavelengths of the lines are in $\mu$m.
}
\medskip
\small
\begin{center}
\begin{tabular}{llcccccl}
\hline\hline
& & & & & & & \\[-7pt]
Star & He\,I & N\,V & \multicolumn{1}{c}{He\,I\,+\,N\,{\sc III}}  & He\,{\sc II}\,+\,Br$\gamma$ & He\,{\sc II} & He\,{\sc II}\,/\,Br$\gamma$& Type\\
& 2.058 & 2.110 & 2.115 & 2.165 & 2.189 & \\
& & & & & & &  \\[-10pt]
\hline
& & & & & & &  \\[-7pt]
1  & P\,Cyg &  no & 8.97   & 23.34 & 14.84 & 0.67 & WN7 \\
   &        &     & 97.2   & 68.7  & 60.7  &      & weak\\
   &        &     &        &       &       &      & \\[-8pt]
2  & P\,Cyg &  no & $-$2.6 & 10.56 & 15.36 & 1.45 & WN6 \\
   &        &     & 82     & 74    & 89    &      & weak\\
   &        &     &        &       &       &      & \\[-8pt]
3  & P\,Cyg &0.39 & 0.98   & 6.26  & no    & 0.67 & Ofp/WN \\
   &        &  19 & 30     & 47    &       &      & ?   \\
   &        &     &        &       &       &      & \\[-8pt]
4  & absorb.&  no &   no   & 2.77  & 2.03  & 0.83 & WN7-8 \\
   &  -2.53 &     &        & 45    & 49    &      & weak\\
   &        &     &        &       &       &      & \\[-10pt]
\hline
\end{tabular}
\end{center}
\label{tab2}
\end{table*}

This spectral classification described above gives a rough estimate
of the temperature and luminosity of the stars. The precision of these
estimates is related to the intrinsic variation within the classes.
Unfortunately, the parameters of the WR stars are not truly
homogeneous. Later in the text we use the WR stars as distance
indicators, so having better estimates is crucial.

The limited number of emission lines in the K atmospheric window
makes the line ratios less reliable than a full line modeling. We used
the CMFGEN model (Hillier \& Miller 1998) to obtain synthetic line
profiles. Even though the spectra do not contain a lot of spectral
features we run models with complex atoms including CNO, Si, P, S and
Fe because they have effect on the electron density, temperature, etc.
and therefore - on the properties of the Hydrogen and Helium lines.

The two main lines in our spectra Br$\gamma$ and
He\,{\sc II}\,2.18\,$\mu$m are sensible to both temperature and mass
loss rate. Their ratio is sensitive to the helium abundance. To
measure all parameters simultaneously we calculated a grid of models
spanning a range of temperatures from 35000 to 40000\,K and mass loss
rates from 1$\times$10$^{-6}$ to 1.5$\times$10$^{-5}$\,${\dot{\cal M}/\rm yr}$.
The chemical composition was set to solar except for He/H which was
set to He/H=1.0, in terms of number of atoms.
Initially, the luminosity of every star was set to $\log$\,L/L$_\odot$=6.0
and the terminal velocity to V$_\infty$=1000\,km/s, determined from the
width of the spectral lines. There are no lines with good P Cyg profile
in the K spectrum, so the value of V$_\infty$ should be treated with
caution. We discus the values of the luminosity below.

\begin{figure}
\centering
\includegraphics[width=\columnwidth]{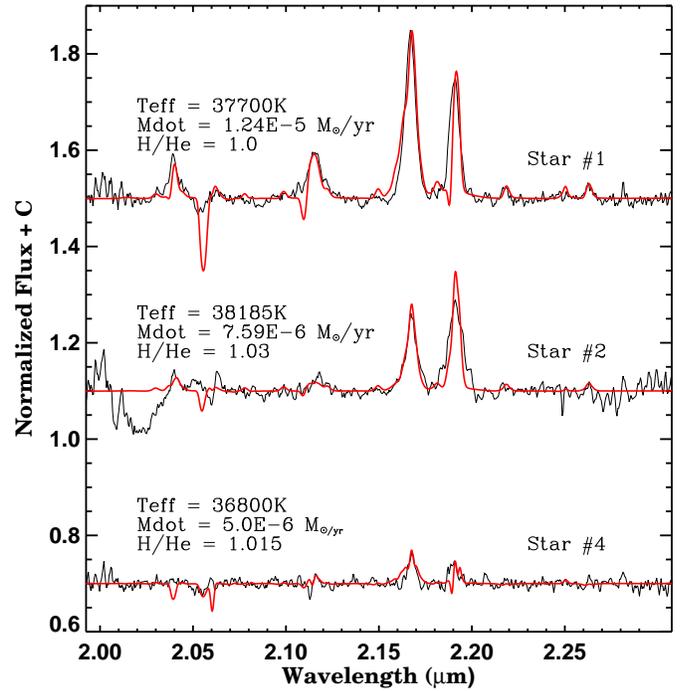}
\caption{CMFGEN model (Hillier \& Miller 1998) fit to the observed
spectra of stars \#1, \#2, and \#4.}
\label{fig7}
\end{figure}

The synthetic spectra were degraded to the resolution of the observed
spectra, using the profile of the arc lamps as a measure of the
instrumental profile (Figure\,\ref{fig7}). Then, we measured the
intensities of Br$\gamma$, He\,{\sc I}\,2.05\,$\mu$m, He\,{\sc I}\,2.11\,$\mu$m
and He\,{\sc II}\,2.18\,$\mu$m on both the models and the observed
spectra. The grid of models defines a surface in T$_{\rm eff}$ vs.
$\dot{\cal M}$ space. The observed value of every spectral feature
is a iso-line on that surface. The crossing of the iso-lines,
determined by different features, sets the parameters which reproduce
simultaneously all measured features. The temperature depends mainly
on the ratio of He\,{\sc I}/He\,{\sc II} lines while the intensity of
Br$\gamma$ is mainly related to $\dot{\cal M}$. Finally, the intensity
of He\,{\sc II}\,2.18\,$\mu$m is sensitive to T$_{\rm eff}$,
$\dot{\cal M}$ and to the helium composition but the first two are
already constrained by the other parameters. We had to adjust the
He/H ratio to reproduce this line well.

\begin{figure}
\centering
\includegraphics[width=\columnwidth]{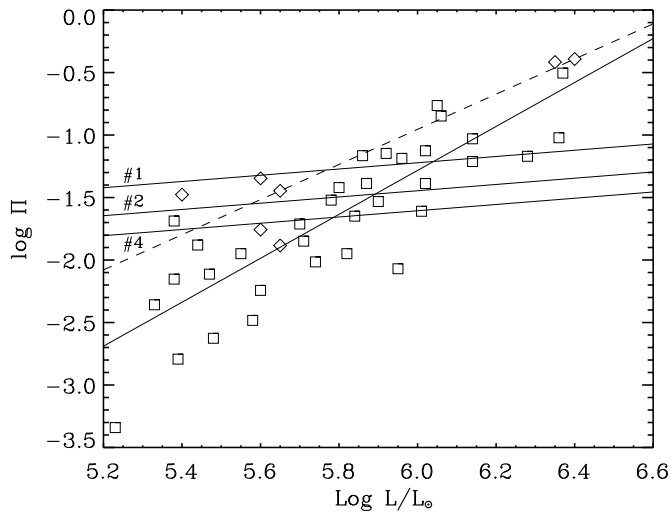}
\caption{Modified wind momentum $\Pi$ versus luminosity $L$ diagram for
a range of radii, for stars \#1, \#2, and \#4. The crosspoint between
the individual line for every star and the observed $\log{L}$$-$$\log{\Pi}$
relation determines the stellar luminosity. WNha stars from Hamann,
Gr\"{a}fener \& Liermann (2006) are marked with diamonds. The least
square fit to these data alone is shown with a dashed line. O stars
from Lamers et al. (1999) are marked with squares. The least square fit
to all data is shown with a solid line.}
\label{fig8}
\end{figure}

As pointed out above, the values of the stellar luminosities were
{\it assumed}. There is no easy way to determine the exact luminosity
from the spectra alone. We broke the luminosity-radius-mass loss rate
degeneracy applying the relation between modified wind momentum
$\Pi$=$\dot{\cal M}$V$_\infty$R$_*^{0.5}$
and the luminosity (Kudritzki, Lenon \& Puls 1995). The derived
T$_{\rm eff}$, $\dot{\cal M}$ and V$_\infty$ set a family of possible
modified wind momenta and stellar luminosities, where the stellar
radius is a free parameter. We calculated the possible $\Pi$ and $L$
for a range of radii and plotted them on the Kudritzki, Lenon \& Puls
(1995) diagram together with data for WNha stars from Hamann,
Gr\"{a}fener \& Liermann (2006). The crossing between this line and
the observed $\log{L}$$-$$\log{\Pi}$ relation determines the stellar
luminosity which gives the same emitted spectrum and satisfies the
$\log{L}$$-$$\log{\Pi}$ relation -- the least square fit to the data
(open diamonds) and is shown in Figure\,\ref{fig8} with a dashed line.
The obtained $\log L/L_\odot$ is 5.77, 5.57 and 5.43 for stars \#1,
\#2, and \#4 respectively. However, the sample of WN stars, containing
hydrogen is very small and we added the data of O-type stars given in
Lamers et al. (1999; shown with squares) to improve the statistics.
The new relation based on the expanded dataset is shown with
continuous line in Figure\,\ref{fig8}. The luminosities are
listed in Table\,\ref{tab3} and we will use them as our final estimates.

This technique allowed us to determine the temperature and the mass
loss rate of the strong He\,{\sc II}\,2.18\,$\mu$m emission stars \#1,
\#2, and \#4. We cross-checked the model predictions with the features
expected in $K$ spectra. The N\,{\sc III} at 2.115\,$\mu$m is heavily
blended with a He\,{\sc I} line which makes the accurate determination
of the nitrogen composition difficult. Nevertheless, the profile fitting
favors an increase of N abundance by a factor of 10 with respect to the
solar. Higher resolution spectra are required to achieve better
precision. The C\,{\sc IV}\,2.07\,$\mu$m is absent from the observed
spectra, which suggests a reduction of the carbon abundance by a factor
of at least 10. In general, these three stars have increased mass loss
rate and relatively low terminal velocity, higher helium abundance and
reduced carbon. All these characteristics make the first two of them
good candidates for hydrogen rich WN stars. The type of the third star
can note be constrained with our data.

\begin{table}
\caption{Physical parameters of WR stars in G30. The bolometric corrections
BC from Crowther et al. (2006), the absolute $K$ magnitude and the true
distance modulus, obtained from WR stars are also listed.}
\medskip
\tabcolsep=3.1pt
\small
\begin{center}
\begin{tabular}{cccccccc}
\hline\hline
& & & & & & & \\[-7pt]
Star & $\log L$ & $T_{\rm eff}$ & $\log \dot{\cal M}$  & H/He & BC & $M_K$ & ($K$-$M_K$)$_0$\\
& [$L_\odot$] & [K] & [${\cal M_\odot}/{\rm yr}$] &  & [mag] & [mag] & \\
& & & & & & &   \\[-10pt]
\hline
& & & & & & &  \\[-7pt]
1  & $6.04$ & $37700$ & $-4.91$   & $1.000$ & $-3.8$ & $-6.55$ &   $15.065$ \\
   &      &           &                   &         &        &         &   \\[-8pt]
2  & $5.91$ & $38185$ & $-5.12$   & $1.030$ & $-4.4$ & $-5.62$ &   $14.134$ \\
   &      &           &                   &         &        &         &   \\[-8pt]
4  & $5.81$ & $36800$ & $-5.30$   & $1.015$ & $-3.6$ & $-6.18$ &   $13.625$\\
   &      &           &                   &         &        &         &   \\[-10pt]
\hline
\end{tabular}
\end{center}
\label{tab3}
\end{table}

\section{Discussion}\label{param}

\subsection{Reddening and Distance}

We used photometric and positional criteria to select probable cluster
members. The candidates occupy a well-defined locus on the color-color
diagram at $H$$-$$K_{\rm S}$$\sim$0.5$-$0.6 \,mag and
$J_{\rm S}$$-$$H$$\sim$0.9$-$1.2 \,mag allowing us to determine the
reddening to the cluster. First, we simply measured the color excesses
of this locus on the color-color diagram with respect to the sequences
of unreddened MS stars (Schmidt-Kaler 1982), obtaining
E($H$$-$$K_{\rm S}$)=0.60$\pm$0.06\,mag and E($J_{\rm S}$$-$$H$)=1.10$\pm$0.06\,mag,
corresponding to $A_{K_{\rm S}}$=1.12$\pm$0.08\,mag and $A_V$=10.00$\pm$0.7\,mag.
Through the paper we used the reddening law of Rieke \& Lebofsky (1985).

The reddening can be determined also from the apparent MS colors, given
that the MS of young clusters is nearly vertical. Applying this method
and fitting only unevolved cluster MS stars with the 4\,Myr Geneva
isochrone (Girardi et al. 2002) we obtained
E($J_{\rm S}$$-$$H$)=1.81$\pm$0.12\,mag, corresponding to
$A_{K_{\rm S}}$=1.19$\pm$0.12\,mag and $A_V$=10.65$\pm$1.1\,mag.

We verified these estimations once again against the spectral type
classification of the WR cluster members, adopting the intrinsic WR
colors of Crowther et al. (2006). The averaged color excesses for the
WN stars are E($H$$-$$K_{\rm S}$)=0.56$\pm$0.07\,mag and
E($J_{\rm S}$$-$$K_{\rm S}$)=1.79$\pm$0.13\,mag, corresponding to
A{$_{K_{\rm S}}$}=1.18$\pm$0.13\,mag and $A_V$=10.54$\pm$1.3\,mag, in
reasonable agreement with our previous estimates.

Ideally, to obtain an accurate distance to the cluster we need the
spectral class and respectively the absolute magnitude of some of the
MS stars because they are much more uniform in comparison with the
WR stars.
Unfortunately, we do not have spectra of MS stars. The WN luminosities
however allow us to obtain relatively good estimate of the distance.
The bolometric corrections (BC) for WN stars from Crowther et al. (2006)
were used to transform the luminosities to absolute $K_{\rm S}$ band
magnitudes. The BCs and the distance moduli of G30 WR stars are listed
in Table\,\ref{tab3}. Averaging the individual estimates, we obtain
A{$_{K_{\rm S}}$}=1.16 and ($K_{\rm S}$$-$$M_{K_{\rm S}}$)$_0$=14.31$\pm$0.35\,mag,
corresponding to distance of d=7.2$\pm$0.9\,kpc.

We also used the mean absolute $K_{\rm S}$-band magnitudes of the WN
subtypes given in Crowther et al. (2006) to verify our distance modulus
estimate. Averaging over the G30 WR members we obtained
($K_{\rm S}$$-$$M_{K_{\rm S}})_0$=13.58$\pm$0.35\,mag. This corresponds
to a shorter distance of d=5.2$\pm$0.9\,kpc.

In our further analysis we will use the larger distance modulus to the
cluster because the models of WR stars yield more reliable temperature
and luminosity estimates (and respectively -- distance) than the spectral
classification based on comparison with template spectra. Furthermore,
the shorter distance modulus moves the hydrogen burning turn-on point,
(where the isochrone of the PMS of 4\,Myr reaches the MS), to about
1\,mag above the observed lower MS cut-off. This is in disagreement
with the estimated age of the cluster (see Figure\,\ref{fig4} and
Subsection\,\ref{age} for details).

\subsection{Cluster Age}\label{age}

It is difficult to obtain a reliable age using only an isochrone MS
fitting of young stellar clusters because of the nearly vertical
linear MS locus. The first constrain comes from the lack of red
supergiants in G30, evident from the CMD (Figure~\ref{fig2}). The
stellar evolutionary models with rotation predict the onset of red
supergiants at $\sim$4.5$-$5\,Myr, defining an upper limit to the
cluster age.

The WR phase is very short lived and the presence of WR stars limits
the maximal age of the cluster (Meynet \& Maeder 2005). All known WR
stars in G30 are of the WN6-7ha (hydrogen rich) subtype. There are
indications that hydrogen rich WN7 stars are descendants of massive
stars with initial masses above 50$-$60$\cal M_\odot$ (e.g., Crowther
et al. 1995). Then an upper age limit of $4-4.5$\,Myr can be set,
independently of the exact metallicity and mass-loss scenario.

The MS turn-off MS point provides a consistency check. Our final
true distance modulus ($K_{\rm S}$$-$$M_{K_{\rm S}})_0$=14.31\,mag
suggests that the brightest unevolved star is O9.5 or B0 setting the
cluster age to $\sim$3$-$4.5\,Myr. This estimate is not independent,
because we used WR stars to obtain the distance to the cluster.

The age of the PMS stars was determined by fitting theoretical 0.1,
1.0, 4.0, 7.0, and 10\,Myr PMS isochrones from Siess, Dufour \&
Forestini (2000) to the CMD (Figure\,\ref{fig4}). Note that the PMS
stars spread over wide age range but the main locus is between the
1 and 4\,Myr isochrones. There are some stars with ages less than
1\,Myr and also a possible concentration of stars around 10\,Myr but
it is too close to the depth of our photometry to draw a certain
conclusion. We refrain from making more accurate conclusion about
the age and the age spread because a significant fraction of the
PMS's falls into the zone of photometric incompleteness. Deeper CMD
is needed for that. However, we point that continuous star formation
scenario or at least an extended burst can not be excluded to have
occurred in G30, based on our data.

The fraction of stars with NIR-excess correlates inversely with
the stellar age, over small age ranges (Hillenbrand 2005). The vast
majority ($\sim$90\%) of stars older than 3-8\,Myr ceases to show
evidence for accretion. The fraction of IR-excess stars in a very
young stellar cluster such as G30 can be used as an age indicator.
We determined that for G30 the fraction is 12\%. The empirical
calibration of Hillenbrand (2005) suggests an age of 3-4\,Myr, in
perfect agreement with our previous estimations.

Similar to Hillenbrand, Bauermeister \& While (2007) we point out
that ages may be affected by the photometric uncertainties and
astrophysical effects such as variability of young objects,
unresolved binaries, etc. In this regard it is important to discuss
the following two questions:

1) How the distance uncertainty affects the PMS-age? -- As discussed
above, we have two distance moduli estimates that differ by
$\approx$\,0.7\,mag. Propagated to the PMS age this corresponds to
age difference of 6-7\,Myr. The shorter distance modulus gives an age
of more than 10\,Myr which is inconsistent with the presence of WNha
stars so the true age uncertainty is even smaller.

2) How the photometric errors affect the age spread? -- The formal
photometric uncertainties at the turn-on point are
$\sigma$$(J_{\rm S}$$-$$K_{\rm S})$=0.12 and
$\sigma$$(K_{\rm S})$=0.10\,mag.
Adding the errors due to the transformations to the standard
$J_{\rm S}HK_{\rm S}$ system, we see that the contribution of the
photometric errors to the age range does not exceed 1-2\,Myr.
Therefore, most of the observed age spread is intrinsic to the PMS
population.

\subsection{Initial Mass Function and Total Cluster Mass}\label{mass}

To calculate the IMF we converted the stellar magnitudes into masses
using the Geneva models. The total mass of the observed cluster
members down to $\sim$2.35\,$\cal M_\odot$ (including the WR stars)
is $\sim$1600\,$\cal M_\odot$ adopting a distance modulus
($K_{\rm S}$$-$$M_{K_{\rm S}}$)$_0$=14.31 and
reddening A{$_{K_{\rm S}}$}=1.16\,mag. The mass estimation of G30
measured from the (($J_{\rm S}$$-$$K_{\rm S})_0$, M$_{K_{\rm S}}$)
diagram as described in Borissova et al. (2003; see their Fig.~8)
gives similar result.

We also constructed a background subtracted mass function of the
cluster and fitted it with a single power-law, obtaining
$\Gamma$=$-1.01$$\pm$$0.03$ (in this terms the Salpeter slope is
$\Gamma$\,$=$\,$-1.35$) over the mass range
$\log \cal M/M_\odot$\,$=$\,$0.75-1.5$ ($5.62$\,$-$\,31.62\,{$\cal M_\odot$}).
The integration over this MF down to $1\,{\cal M_\odot}$ combined
with the masses of the WR stars leads to a total cluster mass of
$\sim$3\,$\times$\,10$^3$\,$\cal M_\odot$. Naturally, this is only a
lower limit.

These results make G30 only two to three times less massive than
some of the most massive young clusters in the Galaxy -- Arches
and Quintuplet.

\subsection{Comparison with Other Massive Clusters and Position in
the Galaxy}

The physical properties of the G30 together with data for Quintuplet,
Arches and the Central cluster from Figer (2004) are summarized in
Table\,\ref{tab4}. Here ${\cal M}_1$ is the total cluster mass in
observed stars, ${\cal M}_2$ is the total cluster mass for all stars
integrated down to 1$\cal M_\odot$, assuming a Salpeter IMF. The
``Radius'' is the radius in parsecs, from the central stellar surface
density peak. The mass densities $\rho_1$ and $\rho_2$ are simply
${\cal M}_1$ and ${\cal M}_2$ divided by the cluster volume. The
``Age'' and ``Luminosity'' are the estimated cluster age and total
luminosity.

Quintuplet shows very similar characteristics to G30. They both have
bigger radius and lower stellar density in comparison with Arches and
Central cluster but Quintuplet is a factor two more massive and a
factor of three more luminous than G30. The Quintuplet is also older
than G30 because it contains one RSG. The properties of G30,
especially the presence of WR stars, make G30 a smaller analog of
Arches, Westerlund\,1 and Quintuplet.

\begin{table*}
\caption{Parameters of G30 compared to the massive clusters in the
Galactic Center (see Section~\ref{mass} for more details. The data
for Quintuplet, Arches and the Central cluster are from  Figer
(2004; see their Table\,1). Note that the total mass and the
corresponding density of G30 are only lower limits.}
\medskip
\tabcolsep=3.1pt
\small
\begin{center}
\begin{tabular}{lcclcccc}
\hline\hline
& & & & & & &\\[-7pt]
Cluster & log(${\cal M}_1$) & log(${\cal M}_2$) & Radius & log($\rho_1$)             & log($\rho_2$)              & Age & log(L)    \\
        & $\cal M_\odot$    & $\cal M_\odot$    & pc     & $\cal M_\odot$\,pc$^{-3}$ & $\cal M_\odot$\,pc$^{-3}$  & Myr & $L_\odot$ \\
\hline
& & & & & & &   \\[-10pt]
G30        & 3.2 & 3.5 & 1.36 & 2.2 & 2.5 & 3-4 & 7.0 \\
Quintuplet & 3.0 & 3.8 & 1.0  & 2.4 & 3.2 & 3-6 & 7.5 \\
Arches     & 4.1 & 4.1 & 0.19 & 5.6 & 5.6 & 2-3 & 8.0 \\
Center     & 3.0 & 4.0 & 0.23 & 4.6 & 5.6 & 3-7 & 7.3 \\
& & & & & & & \\[-7pt]
\hline
\end{tabular}
\end{center}
\label{tab4}
\end{table*}

\begin{figure}
\centering
\includegraphics[width=\columnwidth]{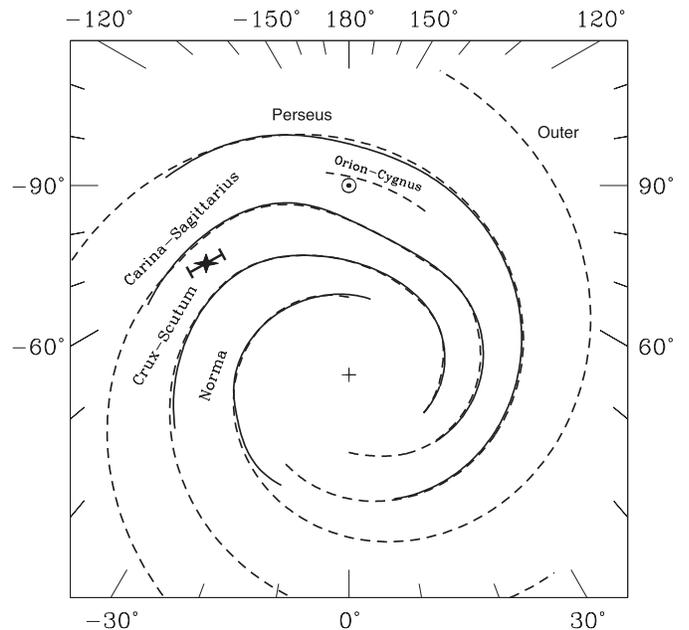}
\caption{Location of G30 in the Galaxy. The galactocentric distance of
the cluster is R$_{GC}$=8.1$\pm$0.4 kpc. The error bar shows the
1$\sigma$ error of 0.9\,kpc in the distance to the Sun.
 The spiral structure is taken from Cordes \& Lazio (2002).
The solid lines show the modified spiral model of the Galaxy from Taylor \&
Cordes (1993).
The dashed lines are a four-arm logarithmic spiral model combined with a
local arm.
The ``+'' sign marks the Galactic center and the Sun is labeled.}
\label{fig9}
\end{figure}

The galactocentric distance of G30 is $R_{\rm GC}$=8.1\,kpc, assuming
that the Sun is at $R_{\rm GC,\odot}$=8.4\,kpc from the Galactic
Center (Figure\,\ref{fig9}).
G30 is located between Carina and Crux spiral arms, closer to Carina.
It probably belongs to some lateral branch of the farther inner side
of this arm. We see the cluster through a dust window in the nearer
side of the Carina arm. Note that if the distance to the cluster is
underestimated it can belong to the farther side of the main Carina
arm.

\section{Summary}

We report new results of our long term project to study obscured Milky
Way star clusters. We obtained deep $J_{\rm S}HK_{\rm S}$ imaging of
G30, a dense stellar cluster with spectroscopically confirmed WR stars
and a sizable population of young stellar objects. G30 is deeply embedded
into gas and dust and suffers a reddening of $A_V$\,$\sim$\,10.5\,mag. The
object probably belongs to the Carina spiral arm and is located at a
distance of 7.2$\pm$0.9\,kpc from the Sun. The cluster is approximately
4\,Myr old. The uppermost MS stars have evolved away from the zero-age
MS. G30 is massive, with a lower limit of the total mass of
$3\,10^3$\,$\cal M_\odot$. This estimate includes only stars with masses
above $\sim$1.0\,$\cal M_\odot$.

Spectral analysis and modeling of $K$ spectra for four objects
show that one of these is Ofpe/WN star, two are hydrogen rich WN6-7
stars, and the last is a WN or O-type star, all with progenitor
masses above 60\,$\cal M_\odot$. The CMD suggests that there might be
more WR or O type cluster members and additional observation are planned
address this possibility. G30 is a new member of the exquisite family of
massive young Galactic clusters, hosting WR stars.

\begin{acknowledgements}
This research is partially supported by the Universidad de Valpara\'iso under DIPUV
grant No 36/2006.  The data used in this paper have been obtained with SofI/NTT at the
ESO La Silla Paranal observatory. This publication makes use of data products from
the Two Micron All Sky Survey, which is a joint project of the University of
Massachusetts and the Infrared Processing and Analysis Center/California Institute
of Technology, funded by the National Aeronautics and Space Administration and
the National Science Foundation. This research has made use of the SIMBAD
database, operated at CDS, Strasbourg, France. The authors would like to thank
Donald Figer and Paul Crowther for placing their spectral libraries to our disposal.
The authors gratefully acknowledge very useful comments of an anonymous referee.
\end{acknowledgements}

\end{document}